\documentclass[journal]{IEEEtran}
\usepackage{graphicx}
\usepackage{amssymb}
\usepackage{amsmath}
\usepackage{balance}
\usepackage{braket}
\usepackage{cite}
\usepackage{color}
\usepackage[switch]{lineno}

\begin{document}

\title{CMOS Quantum Computing:\\ Toward A Quantum Computer System-on-Chip}

\author{Reza~Nikandish,~\IEEEmembership{Senior Member,~IEEE,}~Elena~Blokhina,~\IEEEmembership{Senior Member,~IEEE,}~and Robert~Bogdan~Staszewski,~\IEEEmembership{Fellow,~IEEE}
\thanks{The authors are with the School of Electrical and Electronic Engineering, University College Dublin, Ireland (e-mail: nikandish@ucd.ie).}
}

\maketitle

\IEEEpeerreviewmaketitle

\section{Introduction}

Quantum computing is experiencing the transition from a scientific to an engineering field with the promise to revolutionize an extensive range of applications demanding high-performance computing. The major areas include artificial intelligence, autonomous driving, cryptography, drug development, chemistry, and financial optimization. Many implementation approaches have been pursued for quantum computing systems, where currently the main streams can be identified based on superconducting, photonic, trapped-ion, and semiconductor qubits. Semiconductor-based quantum computing, specifically using CMOS technologies, is promising as it provides potential for the integration of qubits with their control and readout circuits on a single chip. This paves the way for the realization of a large-scale quantum computing system with many qubits (e.g., over 1000) for solving practical problems.

Quantum computing, first envisioned by Richard Feynman and Paul Benioff in the 1980s \cite{feynman82, benioff80}, has passed several important milestones to reach the current state of development. The major landmarks include the invention of the Shor's algorithm for prime number factorization and discrete logarithm on a quantum computer \cite{shor94, shor97}, the development of Grover's algorithm for efficient search in large databases \cite{grover96}, the use of semiconductor quantum dots to implement qubits \cite{loss98}, a silicon-based quantum computer architecture \cite{kane98}, the first spin qubit in silicon \cite{pla12}, the first CMOS spin qubit \cite{maurand16}, and the proposal of using cryogenic CMOS circuits for control and readout of qubits \cite{charbon17}, \cite{patra18}. Recently, Google announced it has achieved the milestone of \emph{quantum supremacy} \cite{preskill12}: in 200 seconds, its Sycamore quantum processor completed a task, the equivalent of which would take a state-of-the-art supercomputer much longer to complete \cite{arute19}.

These achievements along with long-term vision for the future of quantum computing have led to growing global interests and increasing amounts of investment by governments, established companies, and start-ups in this field, e.g., the launch of the US National Quantum Initiative \cite{raymer19, monroe19} and the EU Quantum Technologies Flagship Program \cite{eu_qc3}.

Since the first realization of semiconductor qubits using quantum dots \cite{loss98}, many scientific research works have been devoted to improve the quality of these qubits by using different semiconductor materials (e.g., GaAs, SiGe, and Si) and isotopes (e.g., $\rm ^{28}$Si) \cite{zwanenburg13}. Furthermore, several qubit architectures, including spin-1/2 qubit \cite{levy02}, singlet-triplet qubit \cite{koppens05} and its subset hybrid spin/charge qubit \cite{shi12, kim14, kim15-1}, and exchange interaction qubit \cite{DiVincenzo00-2}, have been proposed to improve the performance and simplify the realization of qubits and logic quantum gates. Most of these developments attempt to improve the performance of individual qubits, e.g., their decoherence time, manipulation time, and fidelity. In a large-scale quantum computing system, however, there are other considerations which can be even more critical than the performance of qubits. For instance, quality of the interface between qubits and classical electronic circuits for control and readout is critical in the performance of a \emph{large-scale} quantum computer.

Quantum computers can outperform classic computers by virtue of running quantum algorithms \cite{Montanaro06}. These algorithms are realized using \emph{quantum circuits} which are sequences of elementary quantum gates applied to qubits. Semiconductor qubits and quantum gates have limited coherence time, e.g., due to environmental noise, which is usually much shorter than the time required to execute quantum algorithms. 
Multiple physical qubits can be employed to construct a logical qubit with much higher performance \cite{fowler12}. The associated quantum error correction improves the fault tolerance of quantum algorithms required for large-scale quantum computation. The hardware overhead of redundant qubits is a challenge in the implementation of quantum error correction. In the Noisy Intermediate-Scale Quantum (NISQ) era, quantum computers are expected to be realized using \emph{imperfect} and \emph{limited} number of qubits (e.g., 50--100). However, the size of quantum circuits will be limited by noise in the quantum gates and, as a result, the quantum computers can outperform classic computers only in a few computational tasks \cite{preskill18}.

CMOS technology can provide potential for the implementation of high-quality qubits, as a result of the high silicon purity achieved in advanced processes \cite{Vandersypen17, blokhina20}. Large arrays of qubits can be implemented in a compact chip area using nanometer-scale transistors. This allows the integration of redundant qubits for quantum error correction with target qubits on the same chip. Furthermore, the use of cryogenic CMOS circuits for control and readout of qubits enables the opportunity of integrating the qubits and their interface circuits on a single chip. This perspective, however, entails dealing with numerous new challenges, including the lack of precise cryogenic models of CMOS devices (e.g., transistors, inductors,  capacitors, resistors, interconnects), process variations, the effects of control circuits on qubit performance, crosstalk between multiple paths of RF control signals, and decoherence of qubits arising from the noise of readout circuits. 

In this paper, we present an overview and future perspective of CMOS quantum computing, exploring developed semiconductor qubit structures, quantum gates, as well as control and readout circuits, with a focus on the promises and challenges of CMOS implementation. In Section~\ref{sec:CMOS_Qubits}, we investigate semiconductor qubit structures and quantum gates. The interface of classic and quantum electronics is elaborated in Section~\ref{sec:quantum_classic_interface}, where we discuss qubit control and readout circuits, as well as architectures for large-scale qubit arrays. In Section~\ref{sec:future}, we present future trends in CMOS quantum computing toward the realization of scalable quantum computers.

\IEEEPARstart{}{}

\section{CMOS Qubits}
\label{sec:CMOS_Qubits}

The fundamental building block of a quantum computer is a qubit, operating based on the \emph{superposition} of two basic quantum states. The qubit's state can be expressed in terms of the basic quantum states, $\Ket{0}$ and $\Ket{1}$, as $\Ket{\psi} = \alpha_0 \Ket{0} + \alpha_1 \Ket{1}$, where $\alpha_0$ and $\alpha_1$ are complex valued  coefficients. The measured qubit state is a random outcome with a probability of $|\alpha_0|^2$ for $\Ket{0}$ and $|\alpha_1|^2$ for $\Ket{1}$, with the constraint of $|\alpha_0|^2 + |\alpha_1|^2 = 1$) \cite{nielsen10}.

The second fundamental feature of qubits, first noted by Albert Einstein in 1935 \cite{einstein35}, is quantum \emph{entanglement}. The quantum state of each qubit in a pair or group of qubits cannot be described independently of the state of the others. That is, the physical properties of entangled qubits, e.g., position, spin, polarization, are correlated, and if one is measured, that of the others are also collapsed. Entanglement plays an essential role in quantum computing.

Qubits have other special features that distinguish them from classic bits. In a perfectly isolated qubit, there exists a definite phase relation between different states, and the system is called to be coherent. In practice, however, due to interactions of the qubit with the physical environment, the coherence is lost with time through a process called quantum decoherence.
Moreover, the qubits are fragile and their quantum state is lost upon measurement. Therefore, the measurement is not deterministically repeatable \cite{nielsen10}.

The physical implementation criteria of quantum computers have been laid down by David DiVincenzo in 2000 \cite{DiVincenzo00} as follows:
1) scalable physical system with well-characterized qubits,
2) the ability to initialize the state of the qubits,
3) decoherence times much longer than the gate operation time,
4) a universal set of quantum gates, and
5) a qubit-specific measurement capability.

Solid-state quantum dot qubits can be constructed based on the charge or spin of electrons (or holes). In a charge qubit, the quantum states can be defined based on the position of the charge in a quantum dot, while the spin qubits operate based on the polarization of the electron's spin. Several physical implementations have been proposed for these qubits, each featuring different benefits and challenges in terms of operation time, decoherence time, control, and readout requirements. We discuss these qubit structures with a focus on the CMOS implementation for large-scale quantum computing.

\begin{figure}[tbh]
\centering
\includegraphics[width=\columnwidth]{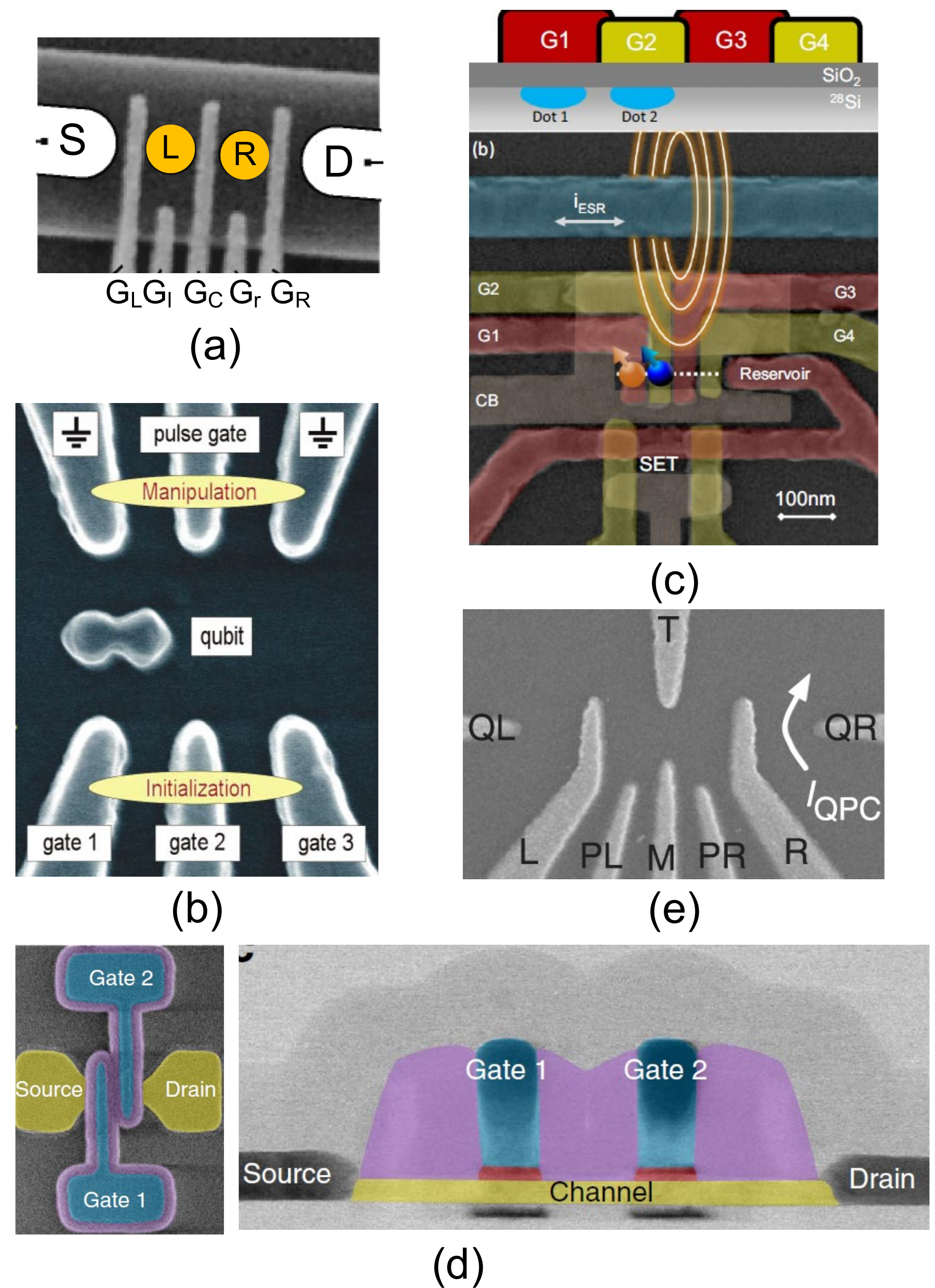}
\caption{Qubit structure candidates for CMOS implementation (a) double quantum dot charge qubit \cite{hayashi03}, (b) isolated double quantum dot charge qubit \cite{gorman05}, (c) MOS spin qubit \cite{hwang17}, (d) SOI CMOS spin qubit \cite{maurand16}, (e) hybrid charge/spin qubit \cite{shi12}.}
\label{cmos_qubits}
\end{figure}

\subsection{Charge Qubits}

The solid-state charge qubits have been extensively studied in the literature \cite{hayashi03, peterson10, petta04, hu05, gorman05, shinkai09, hollenberg04, giounanlis19, blokhina20}. A charge qubit, in its simplest form, can be constructed using a double quantum dot, in which the quantum states are defined by the excess electron occupation in the left and right quantum dots [Fig. \ref{cmos_qubits}(a)]. The two quantum dots are connected through an inter-dot tunneling barrier. The quantum dots are weakly coupled to the source and drain via tunneling barriers. The voltages applied to the left, right, and middle gates control the charge transport through the tunneling barriers. The drain-source voltage can be used to set the operating mode of the qubit. The quantum state of the qubit can be read directly using the drain-source current (for the discussed structure) \cite{hayashi03} or using a charge sensor implemented as a single electron transistor (SET) or quantum point contact (QPC) \cite{peterson10}. This structure should be operated at very low cryogenic temperatures, e.g., $\ll$\,1\,K, to exhibit the quantum effects.
The early solid-state qubits were fabricated using GaAs, while, later, silicon-based structures achieved longer coherence times as a result of specific physical features of silicon \cite{zwanenburg13}.

The charge qubit based on the double quantum dot was first demonstrated in GaAs/AlGaAs hetero-structure and achieved 1\,ns coherence time \cite{hayashi03}. The coherence time improved to 7\,ns using one-electron quantum dots and QPC charge detector \cite{peterson10}. This qubit structure can be made compatible with standard CMOS processes \cite{giounanlis19, blokhina20, bashir20 }. A charge qubit based on an isolated double quantum dot is shown in Fig.\,\ref{cmos_qubits}(b), where the operations are performed using capacitively coupled elements \cite{gorman05}. The quantum state readout is achieved using a SET device integrated with the qubit structure. A prototype implemented in a silicon-on-insulator (SOI) wafer with a phosphorous-doped active region exhibits a long coherence time of 200\,ns. This is attributed to the weak coupling of the isolated qubit to charge noise in the surrounding gates. However, the operational time is relatively short as a result of low inter-dot coupling. This structure requires extra fabrication steps in a conventional CMOS process.
The longest coherence times have been achieved using trapped-ion charge qubits. These structures, however, require extra fabrication steps for precise implantation of donors, which is not fully compatible with current CMOS processes \cite{hollenberg04}.

The charge qubits offer several advantages for CMOS quantum computing. The structure based on the double quantum dot is compatible with standard CMOS processes. Their readout can be performed directly through the drain-source current of the quantum dot or charge sensors integrated with the qubit. Control of the charge qubit can be performed using gate voltage pulses which can be generated with high accuracy in CMOS processes. The quantum gates can be realized as electrostatically coupled quantum dots \cite{shinkai09}. In advanced CMOS, the charge qubits have the potential to achieve fast operating times and maintain coherence significantly longer than the response delay of control and readout circuits. However, there are some challenges that must be addressed for reliable quantum computing using charge qubits. The effects of charge noise on the decoherence and fidelity of charge qubits should be evaluated. Furthermore, the power consumption budget of cooling systems in a large-scale quantum computer limits the minimum temperature of qubits, e.g., 4\,K rather than $\ll$1\,K. This would degrade the performance of the qubits, which is yet to be investigated.

\subsection{Spin Qubits}

The early proposals of silicon quantum computers were based on spin qubits \cite{loss98, kane98}. The solid-state spin qubits can be realized as donor-bound spins or electron spins in quantum dots. The donor-based spin qubits use electrons bound to individual donor atoms, e.g., phosphorous, at cryogenic temperatures. Using these spin qubits in high purity silicon, coherence times exceeding 1\,s are reported \cite{tyryshkin12}, and it is demonstrated that these can be implemented in MOS-like processes with extra fabrication steps \cite{pla12}. The need for precise localization of the donors with respect to electrostatic gates is a major challenge in the use of these qubits for large-scale quantum computing. The spin qubits based on quantum dots should be excited by a magnetic field to control the spin of electrons (or holes).

Several variations of spin qubits based on quantum dots are presented in the literature \cite{levy02, maurand16, hanson07, bluhm19, vandersypen19, hwang17, morton11}. A spin qubit based on MOS double quantum dot structure is proposed in \cite{hwang17} [Fig.\,\ref{cmos_qubits}(c)]. The structure includes four gates (G1--G4) which can be individually tuned to form a quantum dot. A double quantum dot can be created under G1 and G2, which is tunnel-coupled to an electron reservoir under G3 and G4. A SET device is integrated with the structure for charge sensing. The qubit is compatible with standard CMOS processes. Another
CMOS spin qubit is proposed in \cite{maurand16} [Fig.\,\ref{cmos_qubits}(d)]. It is implemented in an SOI process, with two p-channel transistors, one operating as a hole spin qubit, and the other used for the spin readout. There is a buried oxide (BOX) layer between the channel and the silicon substrate, resembling a fully depleted SOI (FDSOI) process. The measured coherence time is 60\,ns. A higher performance is expected by using n-type transistors with electrons as charge carriers. This structure is promising for the realization of scalable quantum computing circuits using a standard FDSOI process. Recently, some spin-based qubit structures are implemented in advanced CMOS and SOI processes \cite{oda16, clarke16, franceschi16, vinet18}, but a standard design procedure and comprehensive characterization are yet to be developed.

\subsection{Hybrid Charge/Spin Qubits}

The hybrid qubit can be viewed as a combination of a spin qubit and a charge qubit \cite{shi12, kim14, kim15-1}. The charge-like characteristics promote a high-speed operation of the qubit, while a long coherence time is achieved due to its spin-like features. The spin operation leads to suppressed charge noise effects because the variations of charge distributions are confined to a single quantum dot \cite{shi12}. The hybrid qubit can be controlled electrically, without the need for magnetic fields required for the manipulation of spin qubits \cite{kim14}. This is an attractive feature for fast operation, scalability, and integration with classical electronics. A possible implementation of this qubit structure is shown in Fig.\,\ref{cmos_qubits}(e), where the gate voltages are used to control the qubit characteristics, and the QPC is used for the quantum state readout. A technique should be developed to control the number of electrons in quantum dots. The hybrid qubit performance and electrical control features are promising for CMOS large-scale quantum computing.

\subsection{Comparison of Qubits for Large-Scale Integration}

It is noted that both the charge and spin qubits can be realized using quantum dot structures in CMOS processes. The choice between two qubits is dependent on several considerations of their performance and interactions with interface circuits. The spin qubits have received more attention because of superior physical properties, mainly the longer coherence time, when considered as \emph{standalone} components. In a \emph{large-scale} quantum computing scenario, however, there are many other considerations which can be even more critical than the coherence time. We briefly discuss the most important aspects.

\begin{enumerate}

\item Coherence time: Spin qubits feature longer coherence time compared to charge qubits. This offers longer time to perform quantum operations. 
\item Operational time: The charge qubits operate faster than the spin qubits. This allows more quantum operations to be completed within the decoherence time of qubits.
\item Sensitivity to charge/spin noise: Semiconductor quantum dots are prone to charge and spin noise, resulting in fluctuating electric and magnetic fields in the qubit, respectively. These effects appear as dephasing and decoherence of the qubit state \cite{kuhlmann13}. The noise can also degrade the fidelity of quantum operations. The charge noise is dominant at low frequencies and follows the $1/f$ spectrum \cite{kuhlmann13, paladino14}, which unfortunately increases at cryogenic temperatures for CMOS transistors \cite{patra18}. The charge qubit features fast operation resulting from strong coupling to electric fields. However, this feature also leads to strong coupling of charge noise which degrades coherence time of the qubit. The spin qubit offers superior noise performance, as a result of the weak coupling of spins to the environment. Some techniques based on symmetric operation and gate pulse engineering are developed to mitigate the qubit noise \cite{reed16, martins16, yang19, kim15}.
\item Quality of the qubit coupling for the realization of quantum gates: The weak interactions of the spin qubits with the environment, which is beneficial to their long coherence times, make inter-qubit operations challenging \cite{schulman12}. The charge qubits, on the other hand, can achieve stronger coupling to each other, enabling the realization of quantum gates through arrays of quantum dots \cite{jones18}.
\item Qubits uniformity in the presence of transistors mismatch: In a large-scale quantum computer with many qubits, mismatch of the qubit characteristics can degrade quality of quantum operations. Qubit structures implemented in advanced CMOS experience greater mismatch at cryogenic temperatures \cite{hart19, hart20}. Furthermore, process-voltage-temperature (PVT) variations can significantly deviate the qubits from their optimal operating conditions. Such issues can be mitigated using digital calibration and error correction techniques in CMOS processes \cite{bashir20, esmailiyan20}.
\item High-temperature operation of qubits: A large-scale quantum computer requires huge power consumption by the dilution refrigerator to maintain qubits at low temperatures essential for their proper operation. This encourages to increase the operational temperature of the qubits, e.g., from $\ll$\,1\,K to 4\,K, to reduce the power consumption as well as size of the cooler. However, this degrades the qubit performance in terms of decoherence, charge noise, and fidelity \cite{petit18}. The effect of increased temperature on performance of the charge and spin qubits as well as quantum gates fabricated using these structures should be evaluated
to ensure feasibility of the high-temperature operation of CMOS quantum computing circuits.
\item Integration of qubits with readout circuits: The charge qubit readout can be performed using charge sensors (e.g., SET or QPC) integrated with the qubit structure. The readout of spin qubits, however, is more complicated and requires the presence of a magnetic field and a spin-to-charge conversion. Therefore, charge qubits are more favorable for a large-scale system.
\item Control circuits: The control of charge qubits can be performed electrically using gate pulses which can be accurately generated by the CMOS circuitry. For the spin qubits, however, a magnetic field is also required. The operating frequency of readout circuits can be as low as $<$\,1\,GHz for the charge qubits while it should be much higher, e.g., 10--20\,GHz, for the spin qubits \cite{bardin20}. A lower frequency of operation alleviates the realization of compact integrated and low-power readout circuits for large-scale qubits. Furthermore, the control circuits introduce external noise into the qubits which can degrade their performance \cite{reilly15, dijk19}. Therefore, sensitivity of spin and charge qubits to such effects should be evaluated for their application in CMOS quantum computing.
\end{enumerate}

\subsection{Quantum Gates}

\begin{figure}[!t]
\centering
\includegraphics[width=\columnwidth]{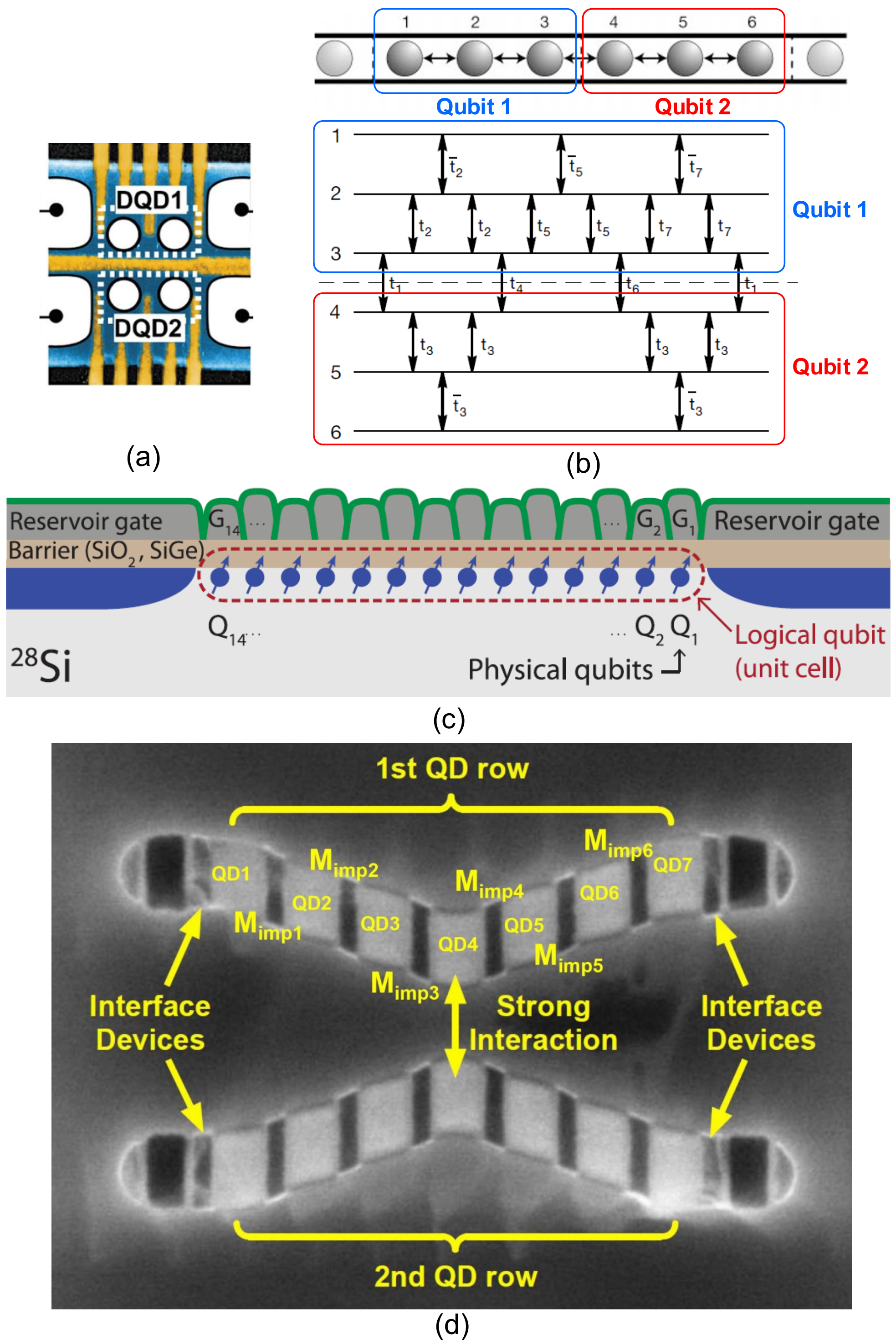}
\caption{Semiconductor quantum gates: (a) electrostatic interaction approach, (b) exchange interaction approach, (c) array of coupled quantum dots, (d) two quantum dot arrays with electrostatic coupling \cite{bashir20}.}
\label{cmos_quantum_gates}
\end{figure}

Quantum gates are operators that evolve the quantum state of qubits \cite{nielsen10, hidary19, sutor19}. Semiconductor-based quantum gates can be realized through interactions between quantum dots \cite{barence95, burkard99, fujisawa11, veldhorst14, Veldhorst15, eenink19, petit20, yang20, zajac16, braakman13, lawrie20, jones18, mills19, bauerle18}. We can consider three main approaches to the physical realization of quantum gates. These methods, shown in Fig.\,\ref{cmos_quantum_gates}, include the electrostatic interaction gates, the exchange interaction between spin-based qubits, and arrays of coupled quantum dots. These can be considered as potential candidates for the CMOS implementation, while still there are many challenges that should be resolved for their reliable operation. 

\subsubsection{Electrostatic Interaction Quantum Gates}

The most straightforward method for the construction of quantum gates is through the electrostatic interaction between quantum dot-based qubits. As shown in Fig.\,\ref{cmos_quantum_gates}(a), two double quantum dots with electrostatic coupling can be used to realize quantum gates \cite{shinkai09, fujisawa11}. In a charge-based operation, coherent oscillations in one double quantum dot are strongly influenced by electron position in the other double quantum dot. The two electrons can simultaneously tunnel in a correlated manner. These coherent oscillations can be interpreted as two-qubit operations. The quantum gate function can be controlled by the coupling strength which is dependent on the physical spacing of the two structures and the voltages applied to the gate terminals of the qubits. Multiple two-qubit operations including CROT, SWAP, and FLIP have been achieved using this technique \cite{shinkai09, fujisawa11}. This structure is scalable and amenable to CMOS processes. The maximum coupling strength is limited by the minimum spacing rules of the CMOS process.

\subsubsection{Exchange Interaction Quantum Gates}

The exchange interaction (Heisenberg interaction) is a quantum mechanical effect resulting from an overlap between the wave functions of two electrons. This effect is proposed to perform quantum operations by using spin qubits \cite{levy02, DiVincenzo00-2}. Single-qubit gates based on spin qubits require stringent control of the magnetic fields applied to the spin and are very slow. An exchange interaction technique is proposed to realize universal quantum gates \cite{DiVincenzo00-2}. Single- and two-qubit gates are implemented using three-state spin qubits in which their interactions are controlled through the time duration of the pulses [Fig.\,\ref{cmos_quantum_gates}(b)]. This structure requires 3$\times$ more devices and about 10$\times$ more clock cycles, which can be readily accommodated by the current CMOS processes benefiting from the small transistor area and accurate clock generation. This technique is particularly useful for reconfigurable and scalable quantum computation as it permits selective choice of single- and multi-qubit quantum operations by turning on/off the coupling between qubits. Using this approach, CNOT, CZ, and SWAP gates have been implemented based on spin as well as hybrid qubits \cite{shi12, DiVincenzo00-2, veldhorst14, Veldhorst15, eenink19, petit20, yang20}. While the operation of single qubits above 1\,K has been demonstrated, it is more challenging for quantum gates. Recently, two-qubit logic quantum circuits operating at 1.1\,K and 1.5\,K are presented in \cite{petit20, yang20}.

\subsubsection{Array of Coupled Quantum Dots}

An array of coupled quantum dots [Fig.\,\ref{cmos_quantum_gates}(c)] can be used to implement various quantum gates \cite{giounanlis19, blokhina20, zajac16, braakman13, lawrie20}. Some of the quantum dots can be exploited as charge injectors or charge sensors. A major challenge is the tuning of the large quantum dot arrays. This structure is amenable to CMOS technology and lends itself to the large-scale integration. The coupled quantum dots can be roughly realized as transistors with a shared drain/source terminal. However, as fewer gate terminals are used for each quantum dot compared to standard structures, it can be difficult to control all features of the quantum dots, e.g., the number of electrons in the dots, the dot potentials, and the width of the depletion layer. The required gate pulse voltages can be generated with high accuracy using CMOS circuits. The quantum dot arrays have been recently implemented in standard CMOS processes \cite{gong19, guevel20, bashir20}, but a detailed qubit characterization has not been reported. This can be an important research endeavor in the future. The quantum dot arrays can also be realized using electrically controllable exchange interaction between spins in adjacent quantum dots \cite{jones18, mills19}. Structures based on two-dimensional quantum dot arrays are developed based on this approach \cite{mukhopadhyay18, mortemousque18, riggelen20}, which can be used in the realization of more complex quantum operations. In \cite{bashir20}, two quantum dot arrays with electrostatic coupling at the middle are used to realize quantum gates [Fig.\,\ref{cmos_quantum_gates}(d)]. Each row is constructed by transistor-like devices operating as quantum dots, imposer, and interface devices. The two arrays are used to realize charge qubits, while their entanglement is controlled by the dot-to-dot distance in the interaction area. This structure is implemented using an FDSOI CMOS process.

\section{Quantum-Classic Electronic Interface}
\label{sec:quantum_classic_interface}

The interface between qubits and classical electronics for the control and readout of their states is of critical importance in a large-scale quantum computer. We discuss the qubit control and qubit readout circuits and proceed with interface architectures for large-scale qubit arrays.

\subsection{Qubit Control Circuits}

In a quantum computer, individual gate bias voltages, control pulses, and, in some cases, microwave signals should be routed to every qubit.
The effects of classical electronic control on the qubit operation are investigated in the literature \cite{dijk19, reilly15, hornibrook15}. For a quantum computer with many qubits (e.g., $>$\,100), it is essential to integrate control circuits with qubits. This provides several advantages in terms of reduced number of control signal paths and interconnect density as well as mitigating the latency and synchronization issues of high-frequency clock propagation over long distances. The cryogenic control circuits can be integrated with qubits, which enables a compact CMOS implementation \cite{dijk19, vandersypen17-2, veldhorst17, bardin19, bashir19, bashir20, esmailiyan20, patra20_ISSCC, guevel20}. The qubit operation and fidelity are affected by the inaccuracies of the control signal, including frequency fluctuations, amplitude and phase noise, jitter, bandwidth, and operating speed \cite{dijk19}. Thermal noise generated by the control circuits is reduced by virtue of operating them at cryogenic temperatures. This improves the fidelity of the qubits, but also imposes new challenges in the cryogenic circuit design. 

The low-frequency phase/frequency noise of a signal generator is dominated by flicker ($1/f$) noise of transistors, which increases at cryogenic temperatures \cite{patra18}. Furthermore, any mismatch between frequency of the control signal and Larmor frequency of the qubit degrades its fidelity during the qubit idle operation. The presence of spurious tones leads to unwanted Rabi oscillations and reduces the fidelity \cite{dijk19}. These indicate stringent phase/frequency noise and spectral purity requirements for the control signal source. Using microwave burst pulses with optimized frequency, amplitude, and pulse duration, the charge qubit control can be performed close to its sweet spot energy level, where the energy difference between the qubit states is insensitive to the detuning energy variations \cite{kim15}. This approach requires fine control of the gate voltages of quantum dots (e.g., better than 0.1\,mV). This can be achieved using high-resolution digital-to-analog converters (DAC), but with extra circuit and power consumption overheads for large arrays of qubits.

Another important issue in a many-qubit system arises from the coupling between control signal paths of different qubits, which can transfer a common noise to correlated errors between the qubits. This effect should be properly evaluated and mitigated to maintain the qubit performance. The operation of control circuits should be fast compared to the decoherence time of qubits and quantum gates. Considering the typical decoherence times of qubits, the speed of current CMOS processes can meet this requirement. Furthermore, the magnetic field required for the control of spin qubits can be implemented using an on-chip transmission line excited by a microwave pulse signal [known as an electron spin resonance (ESR) technique]. In a large-scale qubit array, however, this requires extra routing and increases the control complexity. The charge qubits are therefore preferable from the control viewpoint in a large-scale realization.

\subsection{Qubit Readout Circuits}

\begin{figure}[!t]
\centering
\includegraphics[width=\columnwidth]{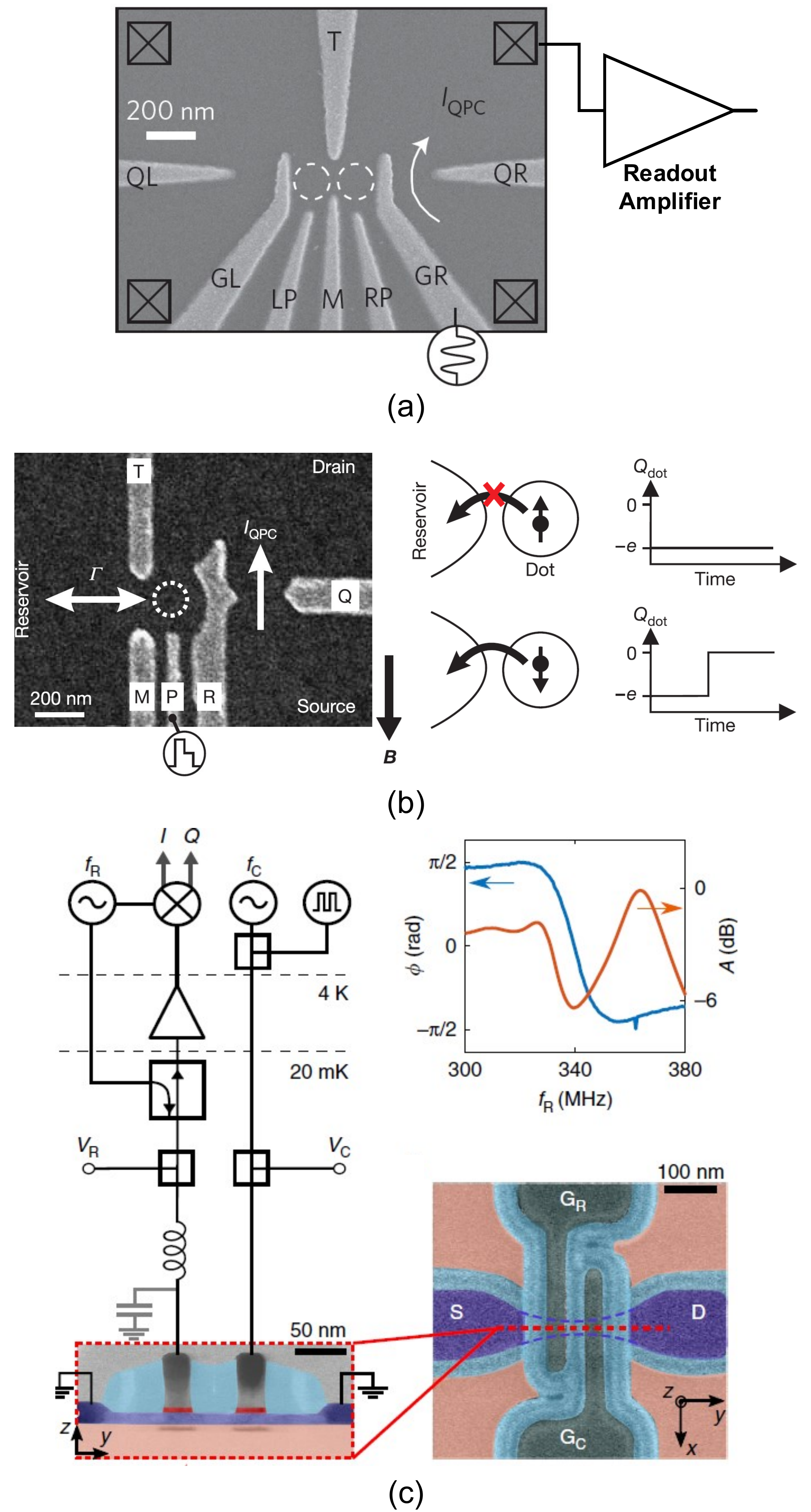}
\caption{The qubit readout techniques: (a) charge sensing \cite{kim15}, (b) spin-to-charge conversion \cite{elzerman04}, (c) RF reflectrometry \cite{crippa19}.}
\label{qubit_readout}
\end{figure}

The readout of qubits is challenging as a consequence of their fragile quantum states. The readout process should be fast, reliable, and scalable so that it can be used in a practical quantum computer. In the case of charge-based qubits, a charge sensor can be integrated with the qubit to detect the presence of an electron in each quantum dot and generate a current which can be measured using a sensitive amplifier [Fig.\,\ref{qubit_readout}(a)]. The readout time is limited by constraints on the minimum noise added by the readout amplifier and a coupling strength between the quantum dot and the charge sensor device. Frequency of the readout circuits for the charge qubits can be low ($<$\,1\,GHz), which allows their readout using mixed-signal circuits with low power consumption and compact chip area in CMOS processes \cite{esmailiyan20}.

Spin states are difficult to read directly, but can be converted to charge states via the spin-to-charge conversion followed by a single-shot charge readout \cite{elzerman04}. As shown in Fig.\,\ref{qubit_readout}(b), a magnetic field is applied to split the two-spin quantum states. The dot potential is tuned such that the electron leaves the dot to the reservoir in the spin-down state, while it stays in the dot in the spin-up state. Therefore, the spin state is correlated with the charge state, which can be read using the charge sensor (QPC here). The spin-to-charge conversion can also be achieved using Pauli spin blockade, which is appropriate for the exchange-interaction spin qubits \cite{petta05}. As the number of qubits is increased, this approach faces challenges due to the required proximity of one charge sensor to each qubit and separate readout amplifier circuits for individual qubits.

RF reflectometry is another readout technique for charge and spin qubits in which a reactive resonator circuit is coupled to the quantum dot to measure changes in its impedance with the quantum state \cite{petersson10, crippa19, west19, schaal19}. The qubit has a state-dependent quantum capacitance which introduces a shift in the frequency response (e.g., phase of the reflection coefficient) of the resonator circuit [Fig.\,\ref{qubit_readout}(c)]. This shift can be measured to detect the quantum state. The resonance circuit and other readout circuitry can be integrated in a CMOS implementation. A highly stable and low-phase-noise reference signal, e.g., generated using direct digital synthesis (DDS), is required to enable accurate measurement of the small phase shift resulting from the capacitance change of quantum dots. This approach is promising for large-scale qubit arrays as a shared readout circuit can be used by multiple qubits. Furthermore, a frequency multiplexing architecture can be realized by using multiple resonator circuits for simultaneous readout of multiple qubits \cite{petersson10}. The frequency of readout circuits is typically in the range of 10--20\,GHz for the spin qubits. An implementation challenge arises from the large size of separate resonator circuits required for each qubit. Furthermore, the parasitic coupling between the frequency-multiplexed qubits can be high at such frequencies.

\subsection{Interface to Large-Scale Qubit Arrays}

\begin{figure}[!t]
\centering
\includegraphics[width=0.9\columnwidth]{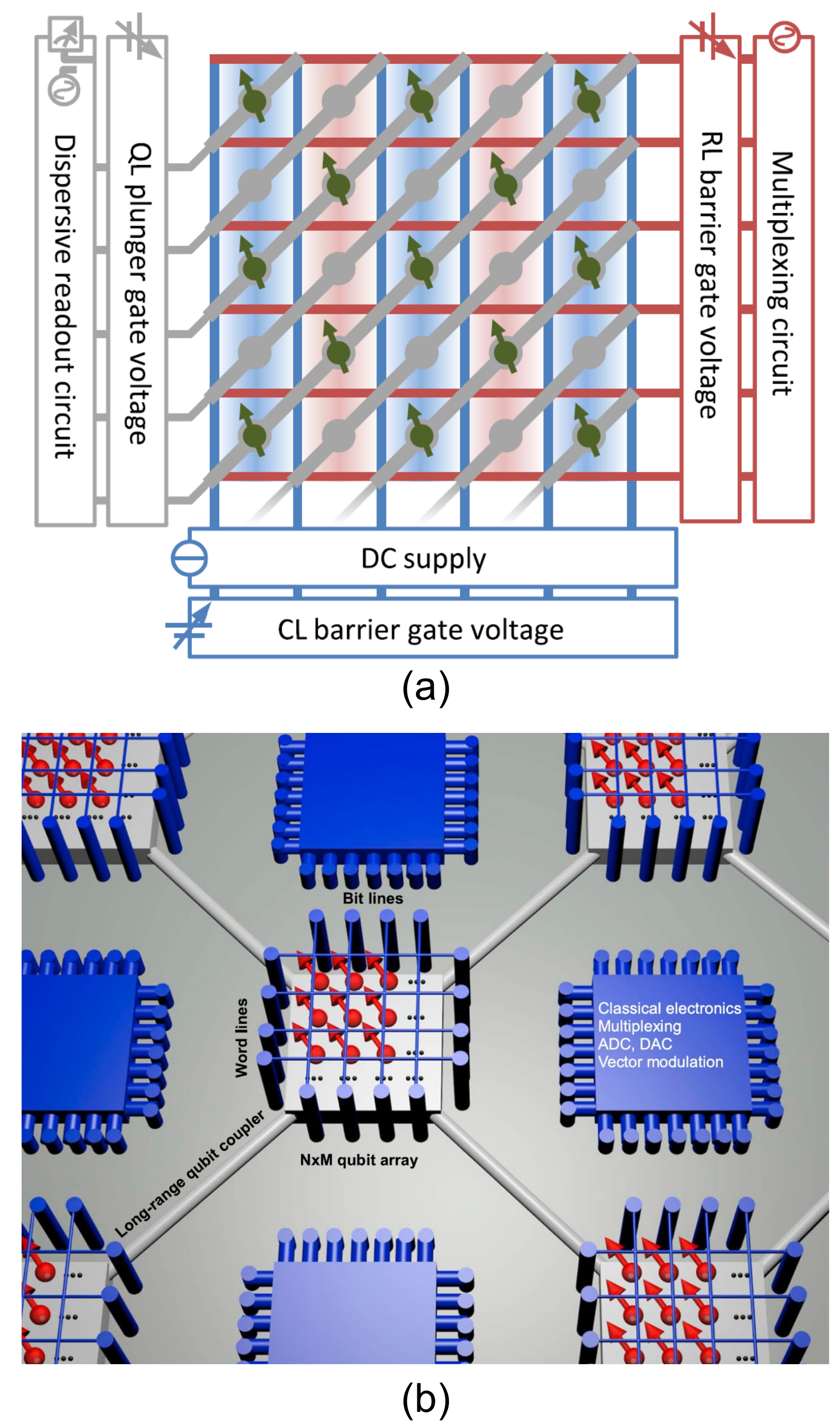}
\caption{Control and readout interface architectures for large-scale qubit arrays: (a) crossbar (DRAM-like) dense qubit array \cite{veldhorst17, li18}, (b) sparse qubit array \cite{vandersypen17-2}.}
\label{qubit_arrays}
\end{figure}

A large-scale quantum computer can be realized using two-dimensional qubit arrays. Such structures can be implemented in CMOS processes with nanoscale feature size, which allows the integration of qubits with their control and readout circuits on a single chip. The number of qubits that can be integrated are limited by the chip footprint and power consumption of qubits. A significant challenge arises from requirements on the routing of control signals and gate bias voltages to the qubits. The architectures shown in Fig.\,\ref{qubit_arrays} are proposed to address the independent control and readout required for every qubit and chip area limitations.

In a dense qubit array, a crossbar (DRAM-like) network shown in Fig.\,\ref{qubit_arrays}(a) can be used to route the control and readout signals to the qubits \cite{veldhorst17, li18, schaal19}. The row lines (RLs) and column lines (CLs) enable the identification of unique points on the grid for read/write operations. The qubit lines (QLs) connect the plunger gates through vias to the qubits. In the structure proposed in \cite{veldhorst17}, comprising 480 qubits, a single floating gate is used to define each qubit and another single floating gate provides exchange coupling between qubits to perform quantum operations. Qubits implemented in CMOS processes are expected to exhibit mismatched features, e.g., Rabi oscillation frequencies, due to process variations. Therefore, considering the low tolerance levels of qubits, their gate bias voltages and control signals must be independently calibrated \cite{vandersypen17-2}. This has been realized using some extra control lines to tune each qubit in the DRAM-like structure.
In the crossbar network presented in \cite{li18}, a spin qubit module is used. It combines a global charge control, local tuning, and electron shuttling between dots with alternating local magnetic fields and global ESR control.

In a sparse qubit array, the qubits are arranged in arrays of smaller size, as shown in Fig.\,\ref{qubit_arrays}(b) \cite{vandersypen17-2, boter19}. This allows the allocation of more space for the routing and control/readout circuitry. The control circuit transistors can be placed on the same layer as the qubit transistors, to be directly integrated with CMOS fabrication. The local control/readout electronics include digital-to-analog and analog-to-digital converters as well as vector modulators. The optimum array size and performance of the interface electronics should be determined based on features of the CMOS process, the number of qubits, control clock frequency, and available power budget.

A fairly large number of cryogenic quantum controller circuits implemented in standard bulk CMOS and FDSOI processes are recently reported in the literature \cite{bardin19, bashir19, bashir20, esmailiyan20, pauka19, vandijk20_TCASI, patra20_ISSCC, vandijk20_JSSC, guevel20, charbon21}. These include a 4--8 GHz controller for transmon qubits in 28-nm CMOS \cite{bardin19}, a 2.4 GHz controller for charge qubits in 22-nm FDSOI \cite{bashir19, bashir20, esmailiyan20}, a controller in 28-nm FDSOI wire-bonded to a 30-qubit GaAs chip \cite{pauka19}, a 2--20 GHz controller for frequency multiplexed spin and transmon qubits in 22-nm FinFET \cite{patra20_ISSCC, vandijk20_JSSC}, and a single chip comprising double quantum dot qubit, 2.8 GHz control circuits, and readout circuits in 28-nm FDSOI \cite{guevel20}.

\section{Future Trends}
\label{sec:future}

The future of quantum computing has been envisioned by many experts representing different viewpoints. We should make a distinction between short- and long-term expectations to distinguish hype from reality \cite{alexeev20}, \cite{corcoles20}. In the NISQ era, the number of qubits (50--100) is less than what can provide a breakthrough in the state-of-the-art computing power \cite{preskill18}. Furthermore, noisy qubits need extensive assistance from the quantum error correction and/or classic machine learning algorithms to mitigate imperfections in the operation of qubits and their control/readout circuitry.

\begin{figure}[!t]
\centering
\includegraphics[width=\columnwidth]{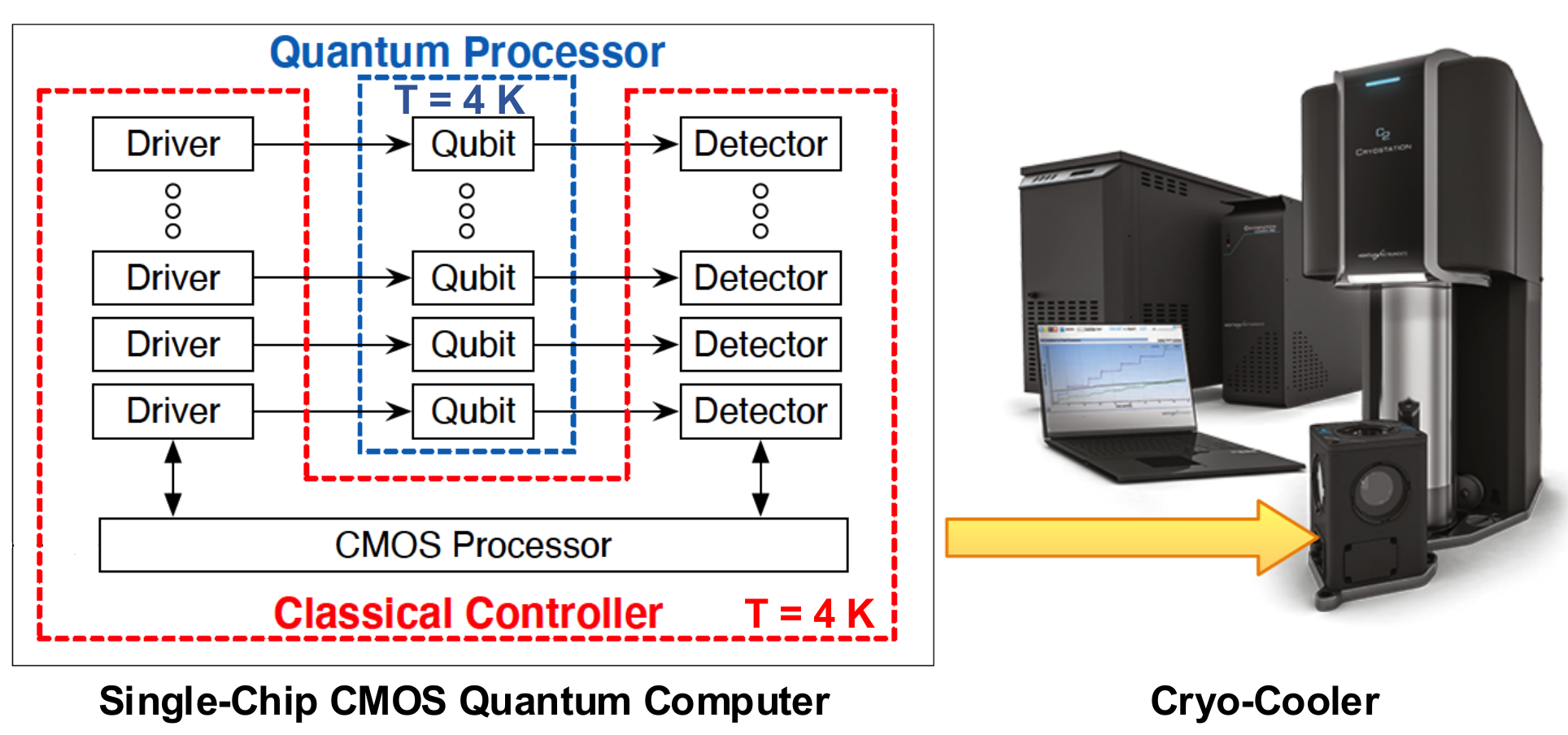}
\caption{Single-chip CMOS quantum processor paradigm where qubits and control/readout circuits are operated at cryogenic temperate of 4\,K to enable their full integration \cite{staszewski20}}
\label{QCCHIP}
\end{figure}

From another perspective, a wide range of application scenarios can be envisioned, from large quantum computers providing cloud-based services to customers, e.g., Google's quantum computing playground and IBM's ``Q Experience", to affordable quantum computers for corporate and personal users. Three levels of development can be envisioned for quantum computing. This starts from the classic scenario in which the quantum processing unit is operating at $<$\,0.1\,K, while the control/readout electronics operates at room temperature. In the next level, the control/readout electronics is moved to the cryogenic temperature of 4\,K. At the ultimate level, both the quantum processing unit and control/readout circuits are operating at the high cryogenic temperature of 4\,K (so-called ``hot qubits"), enabling their full integration in a single chip, as shown in Fig.\,\ref{QCCHIP} \cite{staszewski20}. An important vision is that \emph{current CMOS technologies can provide all the elements required to leverage the realization of personal quantum computers.} CMOS processes provide nanoscale transistors to implement qubits, precise control and readout operations, as well as intensive digital circuits for quantum error correction and calibration \cite{charbon17, Vandersypen17}. One of the key challenges in the realization of quantum error correction is a significant overhead of the required extra qubits ($\sim$100$\times$) \cite{preskill18, hidary19}. This can be feasible using CMOS, as large arrays of qubits can be implemented in a compact chip area. Furthermore, digital calibration techniques can be used to further compensate for the errors.

The implementation of qubits in standard CMOS processes can be an important milestone in quantum computing. Almost all qubits reported so far are fabricated under specially well-controlled lab conditions. To this end, two lines of activity should be pursued. Qubit and quantum gate structures amenable to the mainstream CMOS processes should be developed. This can be complemented by fine-tuning the foundry CMOS processes specifically for quantum applications \cite{govoreanu19}. Furthermore, qubits are usually reported as individual elements with optimized performance (e.g., decoherence time). We emphasize that the definition of high-quality qubit can be different when it should be operated in a \emph{large-scale array integrated with control and readout circuits.} In such operational conditions, there are many other performance metrics which should be considered, including the qubit's interaction with other qubits in the array, resilience against noise, and disturbance of control circuits, coupling to readout circuits, and tolerance of their parasitic elements. This indicates the need for a new design approach for qubit structures amenable to large-scale integration with control and readout circuits.

Recently, several cryogenic CMOS circuits have been presented for quantum computing, including transistor modeling \cite{incandela18, beckers18, bonen19, hart19, hart20, beckers20, yang20_EDL, paz20}, on-chip passive device modeling \cite{patra20_JEDS}, as well as voltage reference \cite{homulle18}, readout amplifiers \cite{gong19}, circulator \cite{ruffino20}, oscillators \cite{gong20}, and a single-electron injection and detection circuits \cite{bashir20, esmailiyan20}. Furthermore, several control/readout circuits have been reported for large-scale qubits \cite{bardin19, bashir19, bashir20-2, pauka19, vandijk20_TCASI, patra20_ISSCC, vandijk20_JSSC, guevel20}. In \cite{bashir20}, charge qubit structures are implemented using a standard 22-nm FDSOI CMOS process. These qubits are integrated with their control and readout circuits \cite{esmailiyan20}. Many auxiliary qubits can be realized on a single chip to perform quantum error correction for the main qubits. This is a promising approach to large-scale quantum computing.

A major challenge in the design of these qubits and their interface circuitry is the lack of accurate cryogenic models for transistors \cite{charbon21}. This has motivated foundries to develop such models. It is expected that foundries will provide cryogenic models in the near future. Another requirement is the development of structures and optimum design approaches (e.g., for low power consumption) for the circuits. Finally, the integration of the CMOS qubit arrays with their control/readout circuitry and mitigating the interface design issues (e.g., noise, parasitic elements, loading effects, undesired coupling) is the last mile toward the realization of an integrated quantum computer system-on-chip (SoC).

\ifCLASSOPTIONcaptionsoff
  \newpage
\fi

\end{document}